\documentclass[aps,pra,twocolumn,groupedaddress,amsmath,amssymb,10pt]{revtex4}
\usepackage{epsfig}
\usepackage{amssymb}
\usepackage{amsmath}
\usepackage{graphicx}
\usepackage{amsfonts}
\usepackage{epsfig}
\usepackage{revsymb}
\usepackage{bm}
\usepackage{amsmath}
\usepackage{amssymb}
\usepackage{latexsym}
\usepackage{hyperref}
\usepackage{cleveref}

\begin{document}

\title{Spin waves in a superconductor}
\author{Serguei N. Burmistrov}
\affiliation{NRC "Kurchatov Institute'', 123182 Moscow, Russia}

\begin{abstract}
 Spin waves that can propagate in normal and superconducting metals are investigated. Unlike normal metals, the velocity of spin waves becomes temperature-dependent in a superconductor. The low frequency spin waves survive within the narrow region below the superconducting transition temperature.  At low temperatures the high frequency waves alone can propagate with an additional  damping due to pair-breaking. 
 \end{abstract}
\maketitle
\par 
It is well-known that the involvement of exchange antiferromagnetic-sign interaction between 
electrons results in a possibility of propagating the spin density oscillations or spin wave modes 
in the Fermi liquids and normal metals \cite{Landau,Silin,Schultz,Platzman}. The conditions of observing the spin waves in superfluid $^{3}\mathrm{He}$-$^{4}\mathrm{He}$ mixtures are discussed in Refs. \cite{Bashkin,Bashkin1}. The standing spin-wave resonances  are observed in normal 
liquid $^{3}\mathrm{He}$ \cite{Masuhara}.  In Ref.~\cite{Fishman} the effects  are studied of the particle-hole asymmetry terms  on the coupling between collective zero-sound and spin-wave modes within the framework of the Landau-Silin equations. Such coupling has permitted a possibility for a detailed study of spin-wave spectrum in normal $^{3}\mathrm{He}$~\cite{Ketterson}. The spin \textit{p}-wave interaction in the particle-particle channel can produce spin excitonic-like and diffusive modes in the superfluid Fermi systems with singlet pairing \cite{Kolom}. 
\par 
The spin wave propagation in superconductors, which is also possible for the antiferromagnetic sign of the electron-electron exchange interaction, has a number of specific features as compared with the normal metals due to formation of electron Cooper pairs with zero total spin. This can be found both in 
an existence of spin waves as a function of temperature and in the wave dispersion. In 
particular, there appears an additional damping of spin waves at the frequencies 
exceeding the energy necessary for breaking an electron Cooper pair off. Such damping is absent in 
normal metals. 
\par
In the frequency region smaller than the superconducting energy gap the spin wave velocity 
proves to be strongly temperature-dependent. In addition, an existence of spin waves becomes 
possible only near the superconducting transition temperature. This is due to strong energy 
dispersion of electron excitations as a function of momentum in the vicinity of the Fermi surface 
and due to drastic reduction of their number with lowering the temperature. The temperature of vanishing the spin waves depends on the magnitude of the exchange coupling constant. Measuring this temperature may give us an information on the electron-electron exchange coupling constant. For the frequencies larger as compared with the superconducting energy gap $\Delta$, the distinction from the normal metal displays in a small  additional damping due to breaking the electron Cooper pairs. 
\par
Prior to turning to superconductor, we recall the spin wave propagation in a normal metal with 
the exchange constant $\lambda (\xi,\xi ')$ depending on the energy of electrons $\xi$ and $\xi '$ as a piecewise function 
\begin{eqnarray*}
\lambda(\xi,\xi')=\left\{
\begin{array}{ccccc}
\lambda +\lambda_1 & \text{for} & \vert\xi\vert & \text{and} & \vert\xi'\vert <\Theta ,
\\
\lambda_1 & \text{for} & \vert\xi\vert & \text{or} & \vert\xi'\vert >\Theta . 
\end{array}
\right. 
\end{eqnarray*}
Here $\xi =\xi_{\bm{p}}$ and $\xi '=\xi_{\bm{p}'}$ are the electron energies taken from the Fermi energy $\varepsilon_F$. 
\par 
So, we represent the exchange coupling as a sum of two contributions. The first one, described by the effective coupling constant $\lambda$, is essential within the energy region smaller than energy $\Theta$. The second one, described by constant $\lambda_1$, extends beyond energy $\Theta$ to the Fermi energy $\varepsilon _F$. Below we keep inequality $\Theta\lesssim\varepsilon_F$ in mind. Concerning the case of superconductor, we can mention a certain analogy of energy $\Theta$ with the Debye frequency  and, correspondingly, constant $\lambda$ with the phonon-mediated interaction and $\lambda_1$ with the Coulomb repulsion between electrons. We put hereafter $\hbar =1$.
\par
Using a ladder geometric series,  we can readily write an integral equation determining the vertex function $\Gamma (\bm{p},\bm{p}';\bm{k})$ for the effective exchange interaction between electrons as
\begin{multline*}
\Gamma (\bm{p},\bm{p}';\bm{k})=\lambda(\xi_{\bm{p}},\xi_{\bm{p}'})+2T\sum\limits_{\varepsilon_1}\int\frac{d^3p_1}{(2\pi)^3}\lambda(\xi_{\bm{p}},\xi_{\bm{p}_1})
\\ 
\times\mathfrak{G}_{\varepsilon_1}(\bm{p}_1)\mathfrak{G}_{\varepsilon_1-\varepsilon}(\bm{p}_1-\bm{k})\Gamma (\bm{p}_1,\bm{p}';\bm{k}).
\end{multline*}
Here symbol $\mathfrak{G}_{\varepsilon}(\bm{p})$ denotes the Matsubara Green's function. Coefficient  2 in the front of the sum over the odd Matsubara frequencies results from the two spin projections. 
\par 
Next, we seek for the solution of the above equation as a piecewise function in accordance with the coupling constant $\lambda(\xi,\xi')$, i.e.
\begin{eqnarray*}
\Gamma (\bm{p},\bm{p}';\bm{k})=\left\{
\begin{array}{ccccc}
\Gamma _0(\bm{k}) & \text{for} & \vert\xi_{\bm{p}}\vert & \text{and} & \vert\xi_{\bm{p}'}\vert <\Theta ,
\\
\Gamma_1(\bm{k}) & \text{for} & \vert\xi_{\bm{p}}\vert & \text{or} & \vert\xi_{\bm{p}'}\vert >\Theta . 
\end{array}
\right. 
\end{eqnarray*}
Then we perform straightforwardly the analytical continuation of the vertex function from the imaginary Matsubara frequencies to the real ones and arrive at the following system of equations:
\begin{gather}
\Gamma_0(\bm{k})=\lambda +\lambda_1 +(\zeta +\zeta_1)A(\bm{k})\Gamma_0(\bm{k})+\zeta_1B(\bm{k})\Gamma_1(\bm{k}), \nonumber
\\
\Gamma_1(\bm{k})=\lambda_1 +\zeta_1A(\bm{k})\Gamma_0(\bm{k})+\zeta_1B(\bm{k})\Gamma_1(\bm{k})\quad\quad\quad
\label{eq1}
\end{gather}
which determine the vertex functions $\Gamma_0(\bm{k})$ and $\Gamma_1(\bm{k})$. The quantities $A(\bm{k})$ and $B(\bm{k})$, expressed in terms of retarded $G^R(\bm{p},\varepsilon)$ and advanced 
$G^A(\bm{p},\varepsilon)$ Green functions for the normal metal, are given by the following relations: 
\widetext
\begin{multline*} 
A(\bm{k})=\int\frac{d\varepsilon}{4\pi i}\int\frac{d\Omega}{4\pi}\int_{-\Theta}^{\Theta}
d\xi_pG^R(\varepsilon,\bm{p})G^A(\varepsilon-\omega, \bm{p}-\bm{k})\biggl(\tanh\frac{\varepsilon}{2T}-\tanh\frac{\varepsilon-\omega}{2T}\biggr),
\\ 
B(\bm{k})=\int\frac{d\varepsilon}{4\pi i}\int\frac{d\Omega}{4\pi}\biggl(\int_{-\infty}^{-\Theta}+\int_{\Theta}^{\infty}\biggr)d\xi_p\biggl[G^R(\varepsilon,\bm{p})G^R(\varepsilon-\omega,\bm{p}-\bm{k})\tanh\frac{\varepsilon-\omega}{2T}
\\ 
-G^A(\varepsilon,\bm{p})G^A(\varepsilon-\omega,\bm{p}-\bm{k})\tanh\frac{\varepsilon}{2T}+G^R(\varepsilon,\bm{p})G^A(\varepsilon-\omega, \bm{p}-\bm{k})
\biggl(\tanh\frac{\varepsilon}{2T}-\tanh\frac{\varepsilon-\omega}{2T}\biggr)\biggr].
\end{multline*}
\widetext
\twocolumngrid 
The coupling constants $\zeta$ and $\zeta_1$  in Eq.~\eqref{eq1} are the exchange constants $\lambda$ and $\lambda_1$ multiplied by the density of states at the Fermi surface. 
\par 
The system of equations \eqref{eq1} delivers us the equation describing the pole of vertex function 
$\Gamma$ and therefore the spin wave dispersion 
\begin{equation}
\label{eq2}
1-A(\bm{k})\biggl(\zeta +\frac{\zeta_1}{1-\zeta_1B(\bm{k})}\biggr)=0.
\end{equation}
The solutions of this equation in the limiting cases represent the typical equation describing the spin wave dispersion with some effective exchange constant $\zeta_{\text{eff}}$ 
$$
-\frac{1}{\zeta_{\text{eff}}}=1+\frac{\omega}{2vk}\ln\frac{\omega -vk}{\omega +vk}
$$
where $v$ is the Fermi velocity. 
\par 
For $\omega\ll\Theta$, quantity $B(\bm{k})$ is about unity and the coupling constants $\lambda$ and $\lambda_1$ summarize approximately as $\zeta_{\text{eff}}=\zeta +\zeta_1$. This is in contrast to the effective coupling attraction governing the superconducting transition temperature where the logarithmic reduction of the repulsive Coulomb potential takes place. 
\par 
For $\omega\gg\Theta$, we have
$$ 
A(k)=-\frac{\Theta}{2vk}\ln\frac{\omega -vk}{\omega +vk}
$$ 
and two solutions of Eq. \eqref{eq2} are possible. The first one with $\zeta_{\text{eff},\, 1}=\zeta\Theta /\omega$ represents the low frequency damping branch. Its existence is wholly determined by the sign of  exchange constant $\zeta$. The second one refers to $\zeta_{\text{eff},\, 2}=\zeta_1$. The survival of the latter branch is fully specified by the sign of exchange constant $\zeta_1$.  
\begin{figure}[ht]
\begin{center}
\includegraphics[scale=0.35]{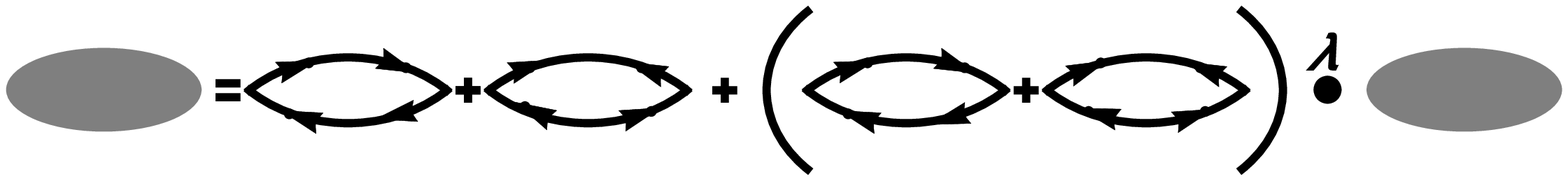}
\caption{The diagrammatic equation for the paramagnetic susceptibility of superconductor. The solid lines are the normal or anomalous Matsubara Green's functions of superconductor. The solid dot denotes  the exchange coupling constant.}
\label{fig1}
\end{center}
\end{figure}
\par 
Let us turn now to the case of superconducting metal and look for the pole of paramagnetic susceptibility. The pole will provide us the dispersion equation for the spin wave propagation. To find the paramagnetic susceptibility of superconductor, we summarize a usual  geometrical ladder series entailing the following diagrammatic representation (Fig.~\ref{fig1}) where $\lambda$ is the exchange constant between the electron spins. 
\par
\begin{figure}[ht]
\begin{center}
\includegraphics[scale=0.35]{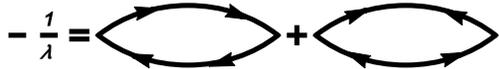}
\caption{The diagrammatic representation of dispersion equation. The solid lines are the normal or anomalous Matsubara Green's functions of superconductor.}
\label{fig2}
\end{center}
\end{figure}

 
The spin wave dispersion equation  
\begin{multline*}
-\frac{1}{\lambda}=2T\sum\limits_{\varepsilon}\int\frac{d^3p}{(2\pi)^3}\biggl[\mathfrak{G}_{\varepsilon_+}(\bm{p}_+)\mathfrak{G}_{\varepsilon_{-}}(\bm{p}_{-})
\\
+\mathfrak{F}_{\varepsilon_+}(\bm{p}_+)\mathfrak{F}^{+}_{\varepsilon_{-}}(\bm{p}_{-})\biggr]
\end{multline*}
expressed in terms of normal $\mathfrak{G}_{\varepsilon}(\bm{p})$ and anomalous $\mathfrak{F}_{\varepsilon}(\bm{p})$ Matsubara Green's functions, can be exemplified by the diagram in Fig. \ref{fig2}. Here momenta stand for $\bm{p}_{\pm}=\bm{p}\pm\bm{k}/2$ and the quantities $\varepsilon_{\pm}=\varepsilon\pm\omega /2$ are the odd Matsubara frequencies. Coefficient  2 in the front of the sum results from the two spin projections. Performing the analytical continuation to the real frequencies, we arrive at the following equation: 
\widetext 
\begin{multline}
-\frac{1}{\lambda}=2\int\frac{d\varepsilon}{4\pi i}\int\frac{d^3p}{(2\pi)^3}\biggl\{\biggl[G^R(\bm{p}_+,\varepsilon_+)G^R(\bm{p}_{-},\varepsilon_{-})+F^R(\bm{p}_+,\varepsilon_+)F^{+R}(\bm{p}_{-},\varepsilon_{-})\biggr]\tanh\frac{\varepsilon-\omega/2}{2T}
\\
-\biggl[G^A(\bm{p}_+,\varepsilon_+)G^A(\bm{p}_{-},\varepsilon_{-})+F^A(\bm{p}_+,\varepsilon_+)F^{+A}(\bm{p}_{-},\varepsilon_{-})\biggr]\tanh\frac{\varepsilon+\omega/2}{2T} 
\\ 
+\biggl[G^R(\bm{p}_+,\varepsilon_+)G^A(\bm{p}_{-},\varepsilon_{-})+F^R(\bm{p}_+,\varepsilon_+)F^{+A}(\bm{p}_{-},\varepsilon_{-})\biggr]\biggl(\tanh\frac{\varepsilon+\omega/2}{2T} -\tanh\frac{\varepsilon-\omega/2}{2T}\biggl)\biggr\}.
\label{eq3}
\end{multline}
\widetext
\twocolumngrid 
This equation corresponds fully to the diagram in Fig.~\eqref{fig2} continued analytically from the imaginary frequencies in the upper half-plane to the real frequency axis. Here $G^R$, $F^R$, and $F^{+R}$ ($G^A$, $F^A$, and $F^{+A}$) are the retarded (advanced) Green functions of superconductor and, correspondingly, analytical in the upper (lower) half-plane. In addition, we have denoted momenta $\bm{p}_{\pm}=\bm{p}\pm\bm{k}/2$ and energies as $\varepsilon_{\pm}=\varepsilon\pm\omega/2$. 
\vspace*{-1em}
\par 
Equation \eqref{eq3} can be simplified as follows:
\begin{multline}
-\frac{1}{\zeta}=\int\frac{d\Omega}{4\pi}\int_0^{\infty}d\xi\frac{\tanh\frac{\varepsilon_+}{2T}-\tanh\frac{\varepsilon_{-}}{2T}}{\varepsilon_{+}-\varepsilon_{-}-\omega -i\delta}
\\
- \int\frac{d\Omega}{4\pi}\int_0^{\infty}d\xi\frac{\varepsilon_+\varepsilon_{-}-\xi_+\xi_{-}-\Delta^2}{\varepsilon_{+}-\varepsilon_{-}-\omega -i\delta} \times 
\\
\biggl[\frac{\tanh\frac{\varepsilon_+}{2T}}{\varepsilon_+(\varepsilon_{+}+\varepsilon_{-}-\omega-i\delta)}-\frac{\tanh\frac{\varepsilon_{-}}{2T}}{\varepsilon_{-}(\varepsilon_{+}+\varepsilon_{-}+\omega+i\delta)}\biggr].
\label{eq4}
\end{multline}
Here $\varepsilon_{\pm}$ denotes $\varepsilon_{\pm}=\sqrt{\xi_{\pm}^2+\Delta^2}$, $\xi_{\pm}=\xi\pm\bm{vk}/2$, $\bm{v}$ is the Fermi velocity and $\Delta=\Delta (T)$ is the superconducting energy gap. The exchange coupling constant $\zeta$ is the coupling constant $\lambda$ multiplied by the density of states at the Fermi surface in the normal state. 
\par
For low frequencies $\omega\sim vk\ll\Delta$, one can readily find from the above equation \eqref{eq4} that the spin wave dispersion is given by the usual relation $\omega =s(T)k$ but with the temperature-dependent velocity $s(T)$. The temperature behavior of velocity $s=s(T)$ should be determined from the following equation:
\begin{eqnarray}
-\frac{1}{\zeta}=\frac{N_n(T)}{N}+\frac{s}{2}\int_0^{\infty}\frac{d\xi}{2T}\frac{1}{\cosh^2(\sqrt{\xi^2+\Delta^2}/2T)} \nonumber
\\
\times\frac{1}{v(\xi)}\ln\frac{s-v(\xi)+i\delta}{s+v(\xi)+i\delta}.
\label{eq5}
\end{eqnarray}
Here $N_n(T)/N$ is the fraction of normal electron excitations in the superconductor. Velocity $v(\xi)$ implies that of excitations in the superconducting state of a metal, i.e. 
$$
v(\xi)=v\frac{\xi}{\sqrt{\xi^2+\Delta^2}}.
$$
\par 
The spin wave ceases its propagation and transforms to the damping diffusive mode at temperature $T_0$ determined by condition $s(T_0)=v$. From equation~\eqref{eq5} we readily find 
$$
\frac{\Delta(T_0)}{\Delta (T=0)}=e^{-1/\zeta}\quad\text{or}\quad\frac{T_c-T_0}{T_c}=\frac{7\zeta (3)e^{-2\gamma}}{8}e^{-2/\zeta} .
\vspace*{1em}
$$
Here $\gamma=0.577\ldots$ is Euler's constant. Thus, as compared with the normal metal, the transition to the superconducting state leads to hindering the spin wave oscillations and limits them to the narrow region below the superconducting transition temperature $T_0$. 
\par 
The involvement of higher order terms in ratio $vk/\Delta$ in Eq.~\eqref{eq4} results in the noticeable  increase of spin wave velocity and in the following dispersion:
$$
\omega =s(T)k\biggl(1+a\frac{\bigl(s(T)k\bigr)^2}{T\Delta (T)}\biggr), 
$$
$a$ being a positive number of order of unity. 
\par 
As it concerns the lower temperatures, the spin wave in superconductor can exclusively propagate provided that $\omega\sim vk\gg\Delta (T)$. In this case the spin wave velocity equals approximately that in the normal state. An additional damping $\gamma (k)$, given by 
$$
\frac{\gamma (k)}{\omega}=\pi\frac{e^{-1/\zeta}}{\zeta}\frac{\Delta (T)}{\omega}\tanh\frac{\omega}{2T}, 
$$ 
is associated with breaking the electron Cooper pairs off. The coupling of spin wave with the diamagnetic current is negligible as well as in the normal metal.

\end{document}